\documentclass[aps,pra,twocolumn]{revtex4}%
\usepackage{amsmath}
\usepackage{amsfonts}
\usepackage{amssymb}
\usepackage{graphicx}%
\providecommand{\U}[1]{\protect\rule{.1in}{.1in}}

\begin{document}
\title{Universal quantum processor already exists and just waits for the proper
programming }
\author{V.M.\ Akulin }
\affiliation{Laboratoire Aime Cotton, Orsay, France}

\begin{abstract}
A possibility of performing the C-NOT gate operation at the ground and the
first excited states of two harmonic oscillators interacting via a two-level
system subject to complete control is demonstrated. The system resembles
Turing machine, where the result of interaction between oscillators and the
two level system is restricted to a certain fixed unitary transformation
matrix, while all the control required for the implementation of the gate is
provided via manipulations with the two-level system, which remains the only
fully-controllable part of the entire system. Each gate operation requires a
"Turing programming", - it can be realized as a series of $\gtrsim63$
elementary unitary operations. The result shows a way how one can construct a
quantum processor in a multimode microwave cavity equipped with a fully
controlled two-level system, such as Josephson junction chip. Parameters of
already existing experimental devises could allow one to perform up to $15$
gate operations in an ensemble of about $10$ qubits.

\end{abstract}
\maketitle

Microwave resonators equipped with a superconducting qubit based on Josephson
junctions\ became one of the most promising experimental tools for quantum
state control \cite{A} and quantum information processing \cite{B}. Though the
Law-Eberly\cite{LE} protocol for the arbitrary quantum state engineering or
certain manipulations with the Schr\"{o}dinger cat states \cite{B} could be
implemented in such machines, strictly speaking, the unitary transformation
group of such systems does not satisfy conditions of the Jurdjevic-Sussman
theorem\cite{S-J} of complete controllability, since it is non-compact
containing as a subgroup the non-compact group $SU(1,1)$ of harmonic oscillator.

However, in the framework of Jaynes-Cummings model, one finds regimes where
the oscillator excitation cannot exceed a trapping level, for which the Rabi
transition frequency to the next photon number state multiplied by the
interaction time equals to a multiple of $2\pi$. In such a regime, a finite
number of lower Fock states get singled out from the entire oscillator Hilbert
spaces and form an isolated Hilbert subspace with a compact group of
transformations, for which the Jurdjevic-Sussman requirements do hold and
hence the complete control of evolution does become achievable. Thought this
control is universal, it is a bit counter-intuitive, since it implies
sequences of $N^{2}$ operation for the dimension $N$ of the Hilbert subspace
with no evident logic of the action at each step - just the overall action of
the sequence yilds the required untary transformation.

Here we show, that in multimode resonator interacting with a single qubit, the
collection of aforesaid subspaces of each mode can form a Hilbert space for
quantum computation. Such a machine is simply a multimode analog of already
existing experimental devices,- it just require a proper programming in the
form of a sequence of\ number of unitary operations per each logic gate
transformation. In a sense, this resembles the Turing machine for the quantum
case, where the logics gates are realized as rather long sequences of basic
machine operations. The question whether or not such an approach based on just
a single two-level system subject to a sequences of operation is preferable as
compared to a system or large number of two-level quantum systems subject to
simpler and intuitively apparent controls\cite{Agraval}\cite{Haroche}%
\cite{CN}\cite{CC}, is a matter of experimental convenience.

Consider in more detail a multimode quantum microwave resonator equipped with
a completely controlled two-level quantum system (qubit), which can interact
with the resonator modes via the "dipole" interaction. First restrict
ourselves to the case of two modes at frequencies $\omega_{1}$ and $\omega
_{2}$. The Hamiltonian of the system reads%
\begin{align}
\widehat{H}  &  =\Delta\widehat{a}\dagger\widehat{a}-\Delta\widehat{b}%
\dagger\widehat{b}+\delta(t)\frac{\widehat{\sigma}_{z}}{2}+X(t)\widehat
{\sigma}_{x}+Y(t)\widehat{\sigma}_{y}\nonumber\\
&  +\left(  \Omega_{1}\widehat{a}\dagger+\Omega_{2}\widehat{b}\dagger\right)
\widehat{\sigma}^{-}+\left(  \Omega_{1}\widehat{a}+\Omega_{2}\widehat
{b}\right)  \widehat{\sigma}^{+}, \label{EQ1}%
\end{align}
where $\widehat{a}\dagger,\widehat{a},\widehat{b}\dagger\widehat{b}$ are the
creation and anihilation operators of the first and the second mode,
respectively, $\widehat{\sigma}_{x}=\widehat{\sigma}^{-}+\widehat{\sigma}^{+}%
$, $\widehat{\sigma}_{y}=i\widehat{\sigma}^{-}-i\widehat{\sigma}^{+}$,
$\widehat{\sigma}_{z}=2\widehat{\sigma}^{-}\widehat{\sigma}^{+}-1$, are Pauli
matrices of the two-level system, $\Delta=\frac{\omega_{1}-\omega_{2}}{2},$
$\delta(t)=\omega(t)-\frac{\omega_{1}+\omega_{2}}{2}$, while $\Omega_{1}$ and
$\Omega_{2}$ are the qubit vacuum Rabi frequencies for the first and the
second mode, respectively. The controlled frequency $\omega(t)$, as well as
"dipole" couplings to controlled "driving fields" $X(t)$ and $Y(t)$ allows one
to exert complete control over the two-level system. We assume%
\begin{equation}
\Delta\gg\Omega_{1},\Omega_{2}. \label{OfResonance}%
\end{equation}

The qubit is set in resonance with modes $1$ and $2$ in alternation, that is
$\delta(t)\equiv\pm\Delta$, for the time intervals $\tau_{1}$ and $\tau_{2}$ ,
respectively, such that
\begin{equation}
\tau_{1}\Omega_{1}\sqrt{2}=\tau_{2}\Omega_{2}\sqrt{2}=2\pi\label{Trap}%
\end{equation}
and thus the Jaynes-Cummings trapping requirement holds for the second excited
states of both mode oscillators. After each of such time intervals the state
of the system experience the evolutions given by the matrices%
\begin{equation}
\widehat{A}=e^{-i\pi\sqrt{2}\left(  \widehat{a}\dagger\widehat{\sigma}%
^{-}+\widehat{a}\widehat{\sigma}^{+}\right)  } \label{EQ3}%
\end{equation}
and%
\begin{equation}
\widehat{B}=e^{-i\pi\sqrt{2}\left(  \widehat{b}\dagger\widehat{\sigma}%
^{-}+\widehat{b}\widehat{\sigma}^{+}\right)  }, \label{EQ4}%
\end{equation}
respectively, while the non-resonant corrections to this expression are small
as $\Omega/2\Delta$. In between of these time intervals, one of three
following arbitrary qubit transformations : either
\begin{equation}
\widehat{U}_{x}=\exp[-i\widehat{\sigma}_{x}\sigma], \label{EQ5}%
\end{equation}
with $\sigma=\int_{0}^{\tau}X(t)dt$, or%
\begin{equation}
\widehat{U}_{y}=\exp[-i\widehat{\sigma}_{y}\sigma], \label{EQ6}%
\end{equation}
with $\sigma=\int_{0}^{\tau}Y(t)dt$, or%
\begin{equation}
\widehat{U}_{z}=\exp[-i\widehat{\sigma}_{z}\sigma], \label{EQ7}%
\end{equation}
with $\sigma=\int_{0}^{\tau}\frac{\delta(t)}{2}dt$ and $\Delta\gg\delta(t)$,
is performed. Here $\tau$ is a duration of the time interval of these
manipulations. During these intervals the oscillators experience the phase
transformations%
\begin{equation}
\widehat{U}_{\tau}=e^{-i\tau\Delta\left(  \widehat{a}\dagger\widehat
{a}-\widehat{b}\dagger\widehat{b}\right)  }. \label{EQ8}%
\end{equation}

As one can check directly, due to the condition Eq.(\ref{Trap}), the
transformations $\widehat{A}$, $\widehat{B}$, $\widehat{U}_{x}$, $\widehat
{U}_{y}$, $\widehat{U}_{z}$, $\widehat{U}_{\tau}$ leave invariant the Hilbert
subspace $\mathcal{H}_{\left\{  0,1\right\}  }$\ spanned by $N=8$ vectors%
\begin{equation}%
\begin{array}
[c]{cccccccc}%
\left\vert 111\right\rangle  & \left\vert 110\right\rangle  & \left\vert
101\right\rangle  & \left\vert 100\right\rangle  & \left\vert 011\right\rangle
& \left\vert 010\right\rangle  & \left\vert 001\right\rangle  & \left\vert
000\right\rangle ,
\end{array}
\label{BASIS}%
\end{equation}
where the first index denotes the ground or the excited states of the qubit
and the second and the third indices denote that for the second and the first
modes, respectively. These unitary operators have the matrix structure
\[
\widehat{A},\cdots\widehat{U}_{\tau}=\left(
\begin{array}
[c]{cc}%
\ldots & \widehat{0}\\
\widehat{0} & 8\times8\quad matrix\quad in\quad\mathcal{H}_{\left\{
0,1\right\}  }%
\end{array}
\right)  ,
\]
and as long as the initial state also belongs to $\mathcal{H}_{\left\{
0,1\right\}  }$, the operations $\widehat{A}$, $\widehat{B}$, $\widehat{U}%
_{x}$, $\widehat{U}_{y}$, $\widehat{U}_{z}$ and their products can be
considered in this basis exclusively. The most important property of these
products, is that they form a compact semigroup which contains all unitary
operations in $\mathcal{H}_{\left\{  0,1\right\}  }$, and therefore allow one
to perform in this subspace any desired unitary transformation in the form of
a finite number of sequential transformations Eqs.(\ref{EQ3}-\ref{EQ8}). The
transformations sequence should only depend on a finite number $M\geq
N^{2}-1=63$ of parameters $\sigma$.

More specifically, for $\Delta\tau=2n\pi$, where $n$ is an integer and hence
$\widehat{U}_{\tau}=\widehat{1}$, a sequence%
\begin{align}
\widehat{U}_{d}  &  =e^{-i\widehat{\sigma}_{z}\sigma_{72}}e^{-i\pi\sqrt
{2}\left(  \widehat{a}\dagger\widehat{\sigma}^{-}+\widehat{a}\widehat{\sigma
}^{+}\right)  }e^{-i\widehat{\sigma}_{y}\sigma_{71}}e^{-i\pi\sqrt{2}\left(
\widehat{b}\dagger\widehat{\sigma}^{-}+\widehat{b}\widehat{\sigma}^{+}\right)
}\nonumber\\
&  e^{-i\widehat{\sigma}_{x}\sigma_{70}}e^{-i\pi\sqrt{2}\left(  \widehat
{a}\dagger\widehat{\sigma}^{-}+\widehat{a}\widehat{\sigma}^{+}\right)  }\ldots
e^{-i\widehat{\sigma}_{z}\sigma_{3}}e^{-i\pi\sqrt{2}\left(  \widehat{a}%
\dagger\widehat{\sigma}^{-}+\widehat{a}\widehat{\sigma}^{+}\right)
}\nonumber\\
&  e^{-i\widehat{\sigma}_{y}\sigma_{2}}e^{-i\pi\sqrt{2}\left(  \widehat
{b}\dagger\widehat{\sigma}^{-}+\widehat{b}\widehat{\sigma}^{+}\right)
}e^{-i\widehat{\sigma}_{x}\sigma_{1}}e^{-i\pi\sqrt{2}\left(  \widehat
{a}\dagger\widehat{\sigma}^{-}+\widehat{a}\widehat{\sigma}^{+}\right)  },
\label{PRO}%
\end{align}
can be set equal to any desired $8\times8$ unitary matrix $\widehat{U}_{d}$ by
a proper choice of the parameters $\left\{  \sigma_{i}\right\}  =\left\{
\sigma_{M=72}\ldots\sigma_{1}\right\}  $. As it has already been mentioned,
there is no evident logic in the choice of the action $\sigma_{i}$ at each
step $i$, only the overall result of the sequence maters. In particular, for
the sequence of $72$ parameters $\left\{  \sigma_{i}\right\}  =\{$-0.2872,
0.1842, -0.5489, 0.2484, 0.0132, -0.1134, -0.5642, -0.5800, -2.4470, -0.0432,
0.3052, 0.0869, 0.5365, 0.6245, -0.7469, -0.5959, -0.9621, -2.0245, -0.0107,
0.3731, 0.0410, 0.3369, 0.4287, 0.1212, -0.6637, -0.1490, -2.5645, -0.0396,
-0.0460, 0.1488, 0.2528, 0.4742, 0.9225, - 0.3419, -0.4538, -3.4287, -0.2260,
0.0561, 0.5803, 0.7112, 0.8276, -0.1700, -0.1722, -0.6864, -2.6273, 0.4602,
0.2338, 0.9878, 0.0751, 0.2090, -0.1949, 0.1052, -0.3791, -2.4825, 0.5824,
0.3608, 0.69429, 1.0914, 0.2271, 0.2274, -0.6667, -0.1907, -3.1813, 0.3526,
-0.3946, 0.2783, 0.6658, 0.0545, -0.4650, 0.0846, -0.1140, -2.8158$\}$ the
product Eq.(\ref{PRO}) written in the subspace basis Eq.(\ref{BASIS}) yields a
transformation matrix which up to a global Berry-phase factor, is the C-NOT
gate
\begin{equation}
\widehat{U}_{d}=\left(
\begin{array}
[c]{cccccccc}%
0 & 1 & 0 & 0 & 0 & 0 & 0 & 0\\
1 & 0 & 0 & 0 & 0 & 0 & 0 & 0\\
0 & 0 & 1 & 0 & 0 & 0 & 0 & 0\\
0 & 0 & 0 & 1 & 0 & 0 & 0 & 0\\
0 & 0 & 0 & 0 & 0 & 1 & 0 & 0\\
0 & 0 & 0 & 0 & 1 & 0 & 0 & 0\\
0 & 0 & 0 & 0 & 0 & 0 & 1 & 0\\
0 & 0 & 0 & 0 & 0 & 0 & 0 & 1
\end{array}
\right)  \label{EQ12}%
\end{equation}
for the photonic part. This means that for an arbitrary initial state of the
two-level system, application of the operation sequence Eq.(\ref{PRO})
corresponds to the transformation%
\begin{equation}
\widehat{U}_{C-NOT}=\left(
\begin{array}
[c]{cccc}%
0 & 1 & 0 & 0\\
1 & 0 & 0 & 0\\
0 & 0 & 1 & 0\\
0 & 0 & 0 & 1
\end{array}
\right)  \label{EQ13}%
\end{equation}
in the Hilbert subspace spanned by the ground and the first excited states of
the photonic modes%
\begin{equation}
\left\{
\begin{array}
[c]{cccc}%
\left\vert 11\right\rangle  & \left\vert 10\right\rangle  & \left\vert
01\right\rangle  & \left\vert 00\right\rangle
\end{array}
\right\}  . \label{EQ15}%
\end{equation}
The transformation Eq.(\ref{EQ12}) being a tensor product of the two-level
system identity matrix and the photonic $\widehat{U}_{C-NOT}$ does not produce
entanglement between the qubit and the cavity as long as the latter remains
restricted to first two levels of each mode.

Many other solutions $\left\{  \sigma_{i}\right\}  $ can be numerically found
with the procedure described in \cite{Harel Akulin}\cite{Brion}, and each of
them suits for implementation of the photonic C-NOT gate at the fundamental
transitions of two modes, while the condition $\Delta\tau=2n\pi$ can be
releases just resulting in changing of the sequence $\left\{  \sigma
_{i}\right\}  $. One should also notice, that the sequence Eq.(\ref{PRO}) can
also be build of only two out of three Pauli matrices $\widehat{\sigma
}_{x,y,z}$.

It is evident, that such transformations could be applied to any other pair of
the resonator modes, provided the condition Eq.(\ref{OfResonance}) holds.
Therefore, one can implement C-NOT gates to any of pair of the mode
fundamental transitions thus realizing a universal quantum processor for
microwave photons.

There is, however, a certain pecularity in such a processor. Typically, the
two-qubit gate operation is considered as an "expensive" one, while the local
operations with qubits are "cheap", that is easy to do. This is not the case
for the microwave photons in the setting under consideration -- the local
operations with photons have also to be performed in the same $8$-dimensional
Hilbert subspace and therefore they also require not less than $63$ control
pulses $\left\{  \sigma_{i}\right\}  $. A practical way of doing such an
operation is to first produce the swap matrix
\begin{equation}
\widehat{U}_{s}=\left(
\begin{array}
[c]{cccccccc}%
1 & 0 & 0 & 0 & 0 & 0 & 0 & 0\\
0 & 0 & 0 & 1 & 0 & 0 & 0 & 0\\
0 & 0 & 1 & 0 & 0 & 0 & 0 & 0\\
0 & 1 & 0 & 0 & 0 & 0 & 0 & 0\\
0 & 0 & 0 & 0 & 0 & 0 & 1 & 0\\
0 & 0 & 0 & 0 & 0 & 1 & 0 & 0\\
0 & 0 & 0 & 0 & 1 & 0 & 0 & 0\\
0 & 0 & 0 & 0 & 0 & 0 & 0 & 1
\end{array}
\right)  \label{EQ16}%
\end{equation}
as a sequence o Eq.(\ref{PRO}) with a standard set $\left\{  \sigma
_{i}\right\}  $ of controls, say $\left\{  \sigma_{i}\right\}  =\{$-0.3520,
0.0423, -0.1621, 0.4462 ,0.4655 ,0.4042, 0.2449, -0.1200, -2.5231, 0.077,
0.2609, 0.7865, -0.1527, 0.2210, -1.0893, 0.0321, 0.2538, -1.7061, -0.0262, 0,
0.2744, -0.2684, 0.5115, 0, 0.7084, 0.0365, -2.2245, 0.5371, 0.4411, 0.6516,
0.7463, 1.2677, -0.4479, -0.4177, -0.3899, 0.3146, 0.1395, -0.3993 ,0.2377,
0.0146, 0.3367, 0.3302, -2.6975, -0.4906, -2.4926, -0.0343, 0.0802, -0.1986,
0.6301, 0.5024, 0.8930, -0.2323, -0.3366, -2.7822, 0.3633, 0.3231, 0.2038,
0.0344, 0.3335, -1.1079, -0.0373, 0.1819, -2.3148, 0.1895, -0.1227, 0.4528, 0,
-0.3426, -0.3362, -0.3346, -0.3548, -2.0647 $\}$, and than produce an
arbitrary local operation with the first mode as%
\begin{equation}
\widehat{U}_{1}(\alpha,\beta,\gamma)=\widehat{U}_{s}e^{-i\left(
\alpha\widehat{\sigma}_{x}+\beta\widehat{\sigma}_{y}+\gamma\widehat{\sigma
}_{z}\right)  }\widehat{U}_{s}\text{.} \label{EQ18}%
\end{equation}
The second mode (and the other modes) can be locally controlled by the analogy.

The most fascinating circumstance is that a device suitable for such
operations has already been constructed in reality. From the detailed
manuscript by K. L. Geerlings\cite{Geerlings} one can learn that cryogenic
$3D$ microwave resonators, which have mode frequencies in the domain of $5-15$
$GHz$ and are{\normalsize \ }equiped with a Josephson-junction qubit, can
possess quality factors $Q\gtrsim10^{5}$ and the fundamental Rabi frequencies
$\Omega\sim100$ $MHz$. This implies that during the coherence time of the
system $T\sim100$ $\mu sec$ a series of at least $15$ gate operations each of
which is a sequence of $72$ unitary transformations can be implemented to set
of $10$ modes at a distance $2\Delta\sim1$ $GHz$ one from the other. Of
course, such a processor cannot yes solve the challenging problems of
practical interest, such as factorization of large numbers, but it is already
powerfull enough for implementation of quantum coding protocols against the
coherence losses.


\end{document}